\documentclass[pdflatex,sn-mathphys-num]{sn-jnl}

\usepackage{graphicx}%
\usepackage{multirow}%
\usepackage{amsmath,amssymb,amsfonts}%
\usepackage{amsthm}%
\usepackage{mathrsfs}%
\usepackage[title]{appendix}%
\usepackage{xcolor}%
\usepackage{textcomp}%
\usepackage{manyfoot}%
\usepackage{booktabs}%
\usepackage{algorithm}%
\usepackage{algorithmicx}%
\usepackage{algpseudocode}%
\usepackage{listings}%
\usepackage{siunitx}
\usepackage{textgreek}
\usepackage[utf8]{inputenc}
\usepackage[section]{placeins}
\usepackage{float}
\usepackage{bookmark}

\theoremstyle{thmstyleone}%

\theoremstyle{thmstyletwo}%

\theoremstyle{thmstylethree}%

\newcommand{\etal}{\mbox{\textit{et al.}\ }} 
\DeclareSIUnit\angstrom{\text {Å}}
\DeclareSIUnit\bar{bar}

\begin{document}

\title[Article Title]{Chiral Spinterfaces as an Overlooked Component of the Chiral-Induced Spin Selectivity Effect}

\author[1,2]{\fnm{Franziska} \sur{Schölzel}}
\author[3]{\fnm{Aybüke} \sur{Gülkaya}}
\author[1,2]{\fnm{Rico} \sur{Ehrler}}
\author[2,4]{\fnm{Dominik} \sur{Hornig}}
\author[1,2]{\fnm{Lokesh} \sur{Rasabathina}}
\author[5]{\fnm{Aleksandra}\sur{Lindner}}
\author[5]{\fnm{Jürgen}\sur{Lindner}}
\author[6]{\fnm{Aleksandr} \sur{Kazimir}}
\author[6]{\fnm{Christina} \sur{Lamers}}
\author[1,2]{\fnm{Dietrich R.T.} \sur{Zahn}}
\author[2,4]{\fnm{Michael} \sur{Mehring}}
\author[1,2,5]{\fnm{Olav} \sur{Hellwig}}
\author*[3]{\fnm{Shuxia} \sur{Tao}}\email{S.X.Tao@Tue.nl}
\author*[1,2]{\fnm{Georgeta} \sur{Salvan}}\email{salvan@physik.tu-chemnitz.de}

\affil[1]{\orgdiv{Institute of Physics}, \orgname{Chemnitz University of Technology}, \orgaddress{\street{Reichenhainerstraße 70}, \city{Chemnitz}, \postcode{D-09126}, \state{Saxony}, \country{Germany}}}

\affil[2]{\orgdiv{Research Center for Materials, Architectures and Integration of Nanomembranes (MAIN)}, \orgname{University of Technology Chemnitz}, \orgaddress{\street{Rosenbergstraße 6}, \city{Chemnitz}, \postcode{D-09126}, \state{Saxony}, \country{Germany}}}

\affil[3]{\orgdiv{Department of Applied Physics and Science Education}, \orgname{Eindhoven University of Technology}, \orgaddress{\street{De Zaale}, \city{Eindhoven}, \postcode{NL-5600}, \state{MB}, \country{Netherlands}}}

\affil[4]{\orgdiv{Institute for Chemistry}, \orgname{University of Technology Chemnitz}, \orgaddress{\street{Straße der Nationen 62}, \city{Chemnitz}, \postcode{D-09111}, \state{Saxony}, \country{Germany}}}

\affil[5]{\orgdiv{Institute of Ion Beam Physics and Materials Research}, \orgname{Helmholtz-Zentrum Dresden-Rossendorf}, \orgaddress{\street{Bautzner Landstraße 400}, \city{Dresden}, \postcode{D-01328}, \state{Saxony}, \country{Germany}}}

\affil[6]{\orgdiv{Institute for Drug Discovery}, \orgname{Leipzig University}, \orgaddress{\street{Brüderstraße 34}, \city{Leipzig}, \postcode{D-04103}, \state{Saxony}, \country{Germany}}}

\abstract{
The chiral-induced spin selectivity (CISS) effect is commonly associated with spin-selective transport through chiral molecules, while the role of molecule–electrode interfaces remains poorly understood. Here, we demonstrate that adsorption of chiral amino acid derived molecules on ferromagnetic Ni films with native NiO layer induces a remanent chirality-dependent magneto-optical response. Systematic control experiments establish that this response originates from the molecule-NiO/Ni interface, rather than the molecular layer or bulk ferromagnet. First-principles calculations reveal preferential carboxylate binding and interfacial charge redistribution extending into the NiO surface layers, demonstrating how molecular adsorption modifies the electronic structure of the oxide. Together, these results provide evidence for the formation of \textbf{chiral spinterfaces}, where molecular chirality and interfacial electronic structure give rise to a distinct magnetic response. Our findings highlight the previously overlooked role of molecule-electrode interfaces in CISS-related systems and identify interfacial electronic and magnetic properties as important considerations for understanding and controlling spin selectivity.}


\maketitle

\section{Introduction}\label{sec1}
Chiral materials exist in two non-super-imposable mirror-image forms, commonly referred to as left- and right-handed enantiomers \cite{moss1996basic}. Even though these structures possess identical chemical compositions, their broken mirror symmetry gives rise to distinct interactions with circularly polarized light, magnetic fields, and spin-polarized electrons \cite{bloom2024chiral}. During the past decade, chirality has emerged as an important design principle in molecular electronics, spintronics, catalysis, and quantum materials, driven largely by the discovery of the chiral-induced spin selectivity (CISS) effect \cite{bloom2024chiral,gupta2024chirality,naaman2020chiral,chiesa2023chirality}. In CISS, electron transport through chiral molecules becomes spin selective without requiring strong intrinsic magnetic order, enabling efficient spin filtering under ambient conditions and opening new opportunities for molecular spintronics \cite{bloom2024chiral}.

Despite extensive experimental and theoretical investigations, the microscopic origin of CISS remains under active debate. Existing models primarily attribute the observed spin selectivity to the intrinsic properties of the chiral molecules, invoking combinations of molecular chirality, spin–orbit coupling, and coherent or incoherent electron transport \cite{evers2022theory}. While these approaches have significantly advanced our understanding of spin-selective transport, they generally treat the molecule as the principal active element of the device. By comparison, the role of the molecule–electrode interface, where electronic coupling, charge redistribution, and orbital hybridization are established, has received considerably less attention \cite{hedegard2026geometrycisseffect}.

Because nearly all CISS experiments rely on contacts between chiral molecules and conductive or magnetic substrates, elucidating the role of the molecule-substrate interface in these systems is essential. In photoelectron spectroscopy, transport measurements, Hall-effect devices, and electrochemical experiments, molecular layers are typically assembled on ferromagnetic metals such as Ni or Co \cite{bloom2024chiral,gupta2024chirality}. To prevent oxidation and promote molecular self-assembly, these substrates are frequently coated with thin Au layers that serve as anchoring surfaces for sulfur-containing molecules \cite{nguyen2024mechanism,hornig2026ligand}. Although Au-capped structures have enabled numerous successful CISS measurements, the Au spacer also modifies the electronic coupling between the molecules and the ferromagnet. Consequently, the extent to which the measured spin polarization reflects intrinsic molecular properties, interfacial electronic structure, or both remains a topic of debate.

Independent evidence that interfaces may play an active role comes from the broader field of spintronics, where adsorption of organic molecules on metallic or metal-oxide surfaces is known to modify interfacial electronic and magnetic properties through charge transfer and orbital hybridization \cite{callsen2013magnetic, baljozovicc2026adsorption}. Such adsorption-induced electronic reconstruction can generate spin-polarized interfaces, commonly referred to as spinterfaces, in which the interfacial electronic structure differs fundamentally from that of either the molecule or the substrate alone. Spinterfaces have been extensively investigated for conventional organic–ferromagnet heterostructures because of their ability to control spin injection and transport \cite{callsen2013magnetic,lin2022enhanced,baljozovicc2026adsorption,mollers2022spin,cinchetti2017activating}. However, whether analogous chiral spinterfaces are formed when chiral molecules adsorb on magnetic substrates — and whether such interfaces contribute to the spin polarization measured in CISS experiments — has yet to be established.

Several experimental observations further motivate this question. In particular, Ben Dor~\etal reported that adsorption of chiral molecules on ferromagnetic substrates induces a remanent magnetic response that depends on molecular handedness, an effect termed magnetization induced by proximity of adsorbed chiral molecules (MIPAC) \cite{ben2017magnetization}. Since its original report, however, this phenomenon has received relatively little attention despite its potential relevance to virtually every CISS experiment involving magnetic electrodes. At the same time, numerous studies have demonstrated enantioselective adsorption of chiral molecules on magnetized substrates \cite{safari2024enantioselective,banerjee2018separation}, further emphasizing that molecule–ferromagnet interfaces can exhibit pronounced chirality-dependent electronic interactions. Together, these observations suggest that the interface itself may constitute an active element in spin-selective phenomena rather than merely providing mechanical support for the molecular assembly.

In this work, we directly investigate the formation of chiral spinterfaces using a simplified model system consisting of epitaxial Ni thin films covered only by their native oxide layer, thereby avoiding additional perturbation introduced by Au capping layers. Circular Dichroism (CD) spectroscopy performed in the remanent state enables direct detection of adsorption-induced interfacial magnetic responses without disturbing the magnetic configuration through an external magnetic field. To minimize chemical complexity, we mainly focus on enantiomerically pure amino acids, the small chiral molecular building blocks of the often employed helical polypeptides. The carboxyl- and sulfur side groups of the chosen amino acids enable the formation of well-defined interfaces with the oxide surface. The simple molecular structure compared to other chiral molecules employed in the field allows the relationship between chemical adsorption and interfacial electronic structure to be investigated with minimal ambiguity.

By combining magneto-optical spectroscopy with first-principles calculations, we show that adsorption of chiral molecules generates a chirality-dependent interfacial magneto-optical response that is localized at the molecule–NiO/Ni interface and exhibits reversible magnetic switching. Density functional theory further reveals that adsorption through the carboxylate group leads to charge redistribution extending beyond the first NiO layer, thereby altering the electronic state of the oxide surface. This provides a plausible pathway for transmitting chiral information from the molecule to the NiO surface. Together, the experimental observations and atomistic calculations support the formation of chiral spinterfaces and demonstrate that molecule–ferromagnet interfaces constitute an overlooked but potentially essential component of the CISS effect.

\section{Results}\label{sec2}
\subsection{Adsorption-Induced Interfacial Magneto-optical Response}
To investigate whether adsorption of chiral molecules generates an active interfacial magnetic response, we combined CD spectroscopy with first-principles calculations on a well-defined molecule–NiO/Ni model system. The experiments establish the existence of an adsorption-induced chirality-dependent magneto-optical response, while the calculations provide microscopic insight into the chemical adsorption process and electronic structure of the interface.

We start with the initial characterization of the Ni films to ensure any experimentally observed changes are adsorption-induced. The corresponding data for the following paragraph can be found in the Supplementary Information in Section 1 (Fig.~S1 - Fig.~S5). Nickel thin films (30\,nm) were deposited by electron-beam evaporation onto optically transparent c-plane sapphire substrates (Fig.~S1(a)), with the film thickness chosen to balance optical transmission and bulk magnetic properties (Fig.~S1(b)). Unlike the Au-capped ferromagnetic substrates commonly employed in CISS studies, this architecture enables direct investigation of the molecule–NiO/Ni interface without introducing an additional metallic spacer layer. X-ray diffraction (XRD) confirms epitaxial growth with Ni(111) orientation perpendicular to the substrate surface (Fig.~S2(a)), while the in-plane XRD pole figure reveals the expected six-fold symmetry associated with twin domains in epitaxial Ni(111) films on Al$_2$O$_3$(0001) (Fig.~S3) \cite{fogarassy2013growth}. X-ray Photoelectron Spectroscopy (XPS) and X-ray reflectivity (XRR) measurements identify an approximately 2\,nm native oxide layer (Fig.~S2(b-c)) consisting of mixed NiO and NiOOH (Fig.~S4(a)), leaving a metallic Ni thickness of (27.3 $\pm$ 1) nm (Fig.~S2(b-c)). Superconducting quantum interference device vibrating-sample magnetometry (SQUID-VSM) further demonstrates a well-defined in-plane (ip) easy axis, while out-of-plane (oop) magnetization requires magnetic fields exceeding 0.6\,T (Fig.~S5). Together, these measurements establish a structurally and magnetically well-defined platform for probing adsorption-induced changes in the (magneto-) optical response.

\begin{figure}[!h]
    \centering
    \includegraphics[width=0.85\linewidth]{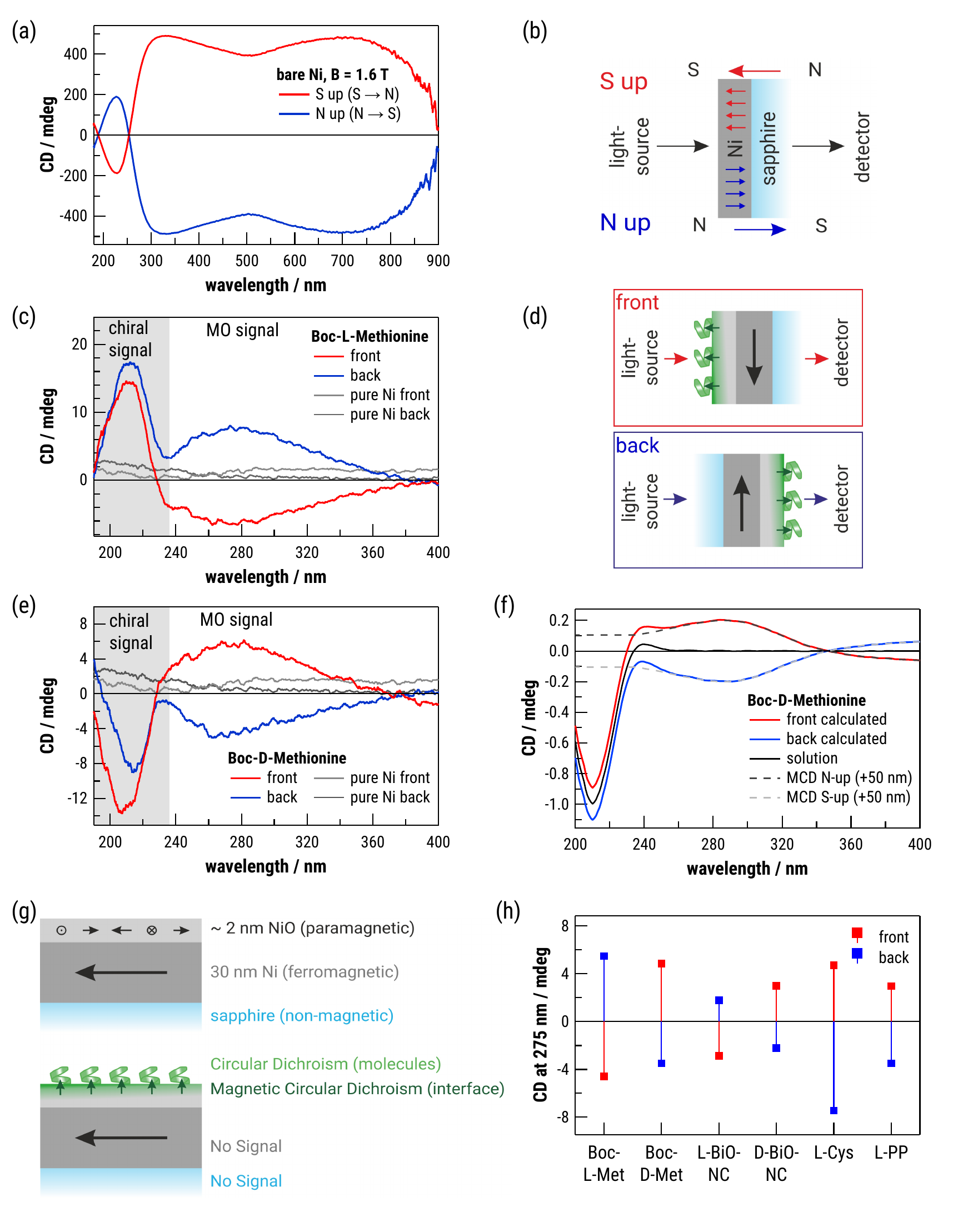}
    \caption{(a) MCD spectra of a bare natively oxidized 30\,nm Ni film with an applied external magnetic field of 1.6\,T in N-up and S-up geometry explained in (b). (b) Scheme of bare Ni film on sapphire in MCD measurement geometry. N-up and S-up denote the two different orientations of the external magnetic field (arrows pointing from N to S) or internal magnetization (blue and red arrows in Ni film). CD spectra in front and back configuration of (c) Boc-L-methionine and (e) Boc-D-methionine on a 30\,nm Nickel thin film on c-plane sapphire. The grey lines correspond to CD measurements of the bare Ni substrate. (d) Illustration of sample orientation of the layer stack in front and back configuration for a left-handed molecule. (f) Calculated lineshape from summation of MCD of bare Ni film (+ 50\,nm) as dotted grey lines and CD spectrum of Boc-D-methionine in solution drawn as black line. (g) Scheme of the layer stack explaining the origin of the adsorption-induced MO signal. The top is showing the bare Ni substrate with layer thicknesses and magnetic properties. The bottom scheme illustrates the sample after adsorption of a left-handed molecule with description of which layers contributing to the observed CD spectrum. (h) CD amplitudes of the adsorption-induced MO signals of various chiral molecules on Ni thin films taken at \SI{275}{nm}. Red squares show the front side values while blue describes the back side.}
    \label{Fig. SPIPAC CD Methionine}
\end{figure}

The intrinsic magneto-optical response of the pristine NiO/Ni substrate was characterized first. Applying an external magnetic field of $\pm$1.6\,T parallel to the propagation direction of the incident light produces a pronounced magnetic circular dichroism (MCD) response (Fig.~\ref{Fig. SPIPAC CD Methionine}(a)), which reverses sign upon field reversal. This behavior reflects the change in the magnetization direction relative to the light propagation direction (see scheme in Fig.~\ref{Fig. SPIPAC CD Methionine}(b)).
\FloatBarrier
In magneto-optical (MO) spectroscopy, active transitions require a change in the total magnetic angular quantum number $m_j$, which is achieved by absorption of circularly polarized photons ($\Delta m_j=\pm1$ for right circular polarized and left circular polarized photons). For comparison, linear polarized light carries $m_j=0$ permitting transitions only between states of identical total angular momentum \cite{mack2007application}. This mechanism gives rise to a chiroptical response even in certain achiral substances exposed to external magnetic fields \cite{djerassi1971organic,kobayashi2011circular}. A similar response is observed whenever a sample possesses a net magnetization component parallel to the light beam (Fig.~\ref{Fig. SPIPAC CD Methionine}(b)). The MCD spectrum of the Ni film (Fig.~\ref{Fig. SPIPAC CD Methionine}(a)) comprises two features of opposite sign: a narrow peak at 220\,nm with an amplitude of approximately 200\,mdeg, and a broad feature extending from 260 to 900\,nm with intensities reaching about 450\,mdeg. The lineshape resembles well the spectra experimentally measured by magneto-optical Kerr effect (MOKE) spectroscopy in this wavelength range as well as the corresponding first principles calculations for interband transitions involving the d-bands \cite{Oppeneer2001MOKE}. We emphasize that the application of an external magnetic field is essential for observing this signal, as the out-of-plane direction corresponds to the hard axis of the Ni film, whereas in the remanent state (in-plane magnetization), no CD signal is detected. These measurements therefore validate the intrinsic magneto-optical fingerprint of the Ni/NiO substrate, against which adsorption-induced changes can be identified.

To make sure that any observed changes in the CD response in remanence were molecule adsorption-induced instead of caused by the cleaning procedure, an identical reference sample was prepared. Here, we followed the same procedure as described in the Methods section but the reference was dipped in spectroscopic-grade ethanol without molecule adsorption. This pristine NiO/Ni substrate exhibits no measurable CD in the remanent state (grey curves in Fig.~\ref{Fig. SPIPAC CD Methionine}(c)), independent of sample orientation (scheme in Fig.~\ref{Fig. SPIPAC CD Methionine}(d)). Additionally, an azimuthal averaging protocol based on the workflow proposed in our recent work was applied, thus making sure that artifacts arising from reflection, linear dichroism, surface roughness, and optical anisotropy are suppressed \cite{schoelzel2026stepbystepworkflowextractgenuine}.

Having now established both a magneto-optical as well as a chiroptical fingerprint of the NiO/Ni substrate, against which changes can be identified, we deposited different chiral molecules via dip coating. Representative CD spectra obtained for the two enantiomers of Boc-methionine (Boc-Met) are shown in Figs.~\ref{Fig. SPIPAC CD Methionine}(c) and (e). The spectra exhibit two distinct features. The first, peaking near 210\,nm, corresponds to the intrinsic chiral signature of Boc-Met \cite{meinert_amino_2022,amdursky_circular_2015,katzin_absorption_1968}. As expected for a molecular chiroptical transition, this band reverses sign between the L- and D-enantiomers while remaining unchanged upon reversing the measurement geometry, where the molecules faced either toward (front) or away from (back) the light source (see Fig.~\ref{Fig. SPIPAC CD Methionine}(d)). This behavior confirms its purely molecular origin.

A second, much broader feature emerges after adsorption of the molecules and is centered near 275\,nm. In striking contrast to the molecular CD band, this feature reverses sign not only between opposite enantiomers but also when the sample is measured in the front and back configurations. Such behavior cannot arise from conventional molecular circular dichroism, which is independent of the direction of light propagation. Additionally, optical artifacts can be excluded due to the applied measurement protocol discussed in our recent work \cite{schoelzel2026stepbystepworkflowextractgenuine}. Instead, the observed reversal is characteristic of a magneto-optical (MO) response associated with a net magnetic moment projected along the optical axis (Fig.~\ref{Fig. SPIPAC CD Methionine}(b)). By adding the normalized, 50\,nm red shifted MCD spectrum of the bare Ni film to the normalized CD measurement of Boc-D-Met in solution, the experimentally observed lineshape from Fig.~\ref{Fig. SPIPAC CD Methionine}(e) is nicely reproduced (Fig.~\ref{Fig. SPIPAC CD Methionine}(f)). From this we assume the molecules induce a net magnetic moment with an oop component during their adsorption on the surface as sketched in Fig.~\ref{Fig. SPIPAC CD Methionine}(g). The opposite responses observed for the two enantiomers must originate from invers directions of the magnetic moments thus directly linking the induced magneto-optical signal to molecular chirality.

To probe how molecular chirality governs the sign of the adsorption-induced MO signal, we examined chiral molecules with diverse architectures but common sulfur-based side groups. The corresponding amplitude of the adsorption-induced CD signal at 275\,nm is displayed in Fig.~\ref{Fig. SPIPAC CD Methionine}(h). Atomically precise bismuth-oxido nanoclusters functionalized with both enantiomers of Boc-Met $\lbrack$Bi$_{38}$O$_{45}$(Boc-Met)$_{24}$$\rbrack$ (L-BiO-NC and D-BiO-NC) \cite{morgenstern2024anchoring,hornig2026ligand} currently shown to exhibit the CISS effect \cite{hornig2026ligand}, generated responses matching the sign of Boc-Met, albeit at reduced amplitude. We attribute this attenuation to lower surface coverage arising from their larger diameter of 2.5\,nm in comparison to the Boc-protected amino acids \cite{morgenstern2024anchoring}. The $\alpha$-L-polypeptide C[AAAAK]$_6$ (L-PP) results in an inverted sign relative to expectation from the notation of its chirality due to the left-handed nature of the amino acids forming the molecule. This discrepancy points to either dominant helicity effects as a right handed helix is formed or the chirality of the adsorbed amino acid is in control. Boc-L-cysteine resolves the ambiguity. Please note that the Fischer and the Cahn-Ingold-Prelog (CIP) assignments agree for most canonical amino acids, L-cysteine is uniquely classified as right-handed under CIP rules \cite{Wolfrom1963,cahn1966specification}. Across all samples, the adsorption-induced MO sign follows absolute structural handedness (CIP) rather than empirical optical activity (Fischer). Given the strongly interfacial nature of the effect, we conclude that the anchoring amino acid and not backbone chirality or overall chirality of the nanoobject determines the signal polarity for the polypeptides. Even though a complete mechanistic account awaits further study, the following sections outline the interfacial character as well as microscopic origin of the MO response to establish a qualitative framework for chiral spinterfaces.

Several observations further demonstrate that the adsorption-induced MO feature does not originate purely from the molecular layer itself. First, the azimuthal averaging protocol effectively suppresses artifacts arising from linear optical anisotropy and reflection \cite{schoelzel2026stepbystepworkflowextractgenuine}. Furthermore, identical measurements performed on 30\,nm Au films after adsorption of the same molecular layers exhibit only the intrinsic molecular CD band and no additional broad feature (Fig.~S6). Thus molecular packing, excitonic coupling, or the molecular film itself are excluded to be the direct source of the signal. Finally, comparison with the saturated out-of-plane MCD response of the pristine Ni film shows that the adsorption-induced MO feature corresponds to only 1–2\,\% of the full MCD intensity, indicating that only a thin region near the surface contributes to the observed response.

Collectively, these findings demonstrate the existence of a previously unrecognized chirality-dependent interfacial magneto-optical response generated by molecular adsorption. The small signal amplitude, together with the control experiments, indicates that the response originates from the molecule–NiO/Ni interface rather than from the molecular layer itself. The following experiments therefore investigate the spatial localization, magnetic character, and microscopic origin of this adsorption-induced interfacial state.
\FloatBarrier
\subsection{Establishing the Interfacial Magnetic State}
\begin{figure}[!ht]
    \centering
    \includegraphics[width=\linewidth]{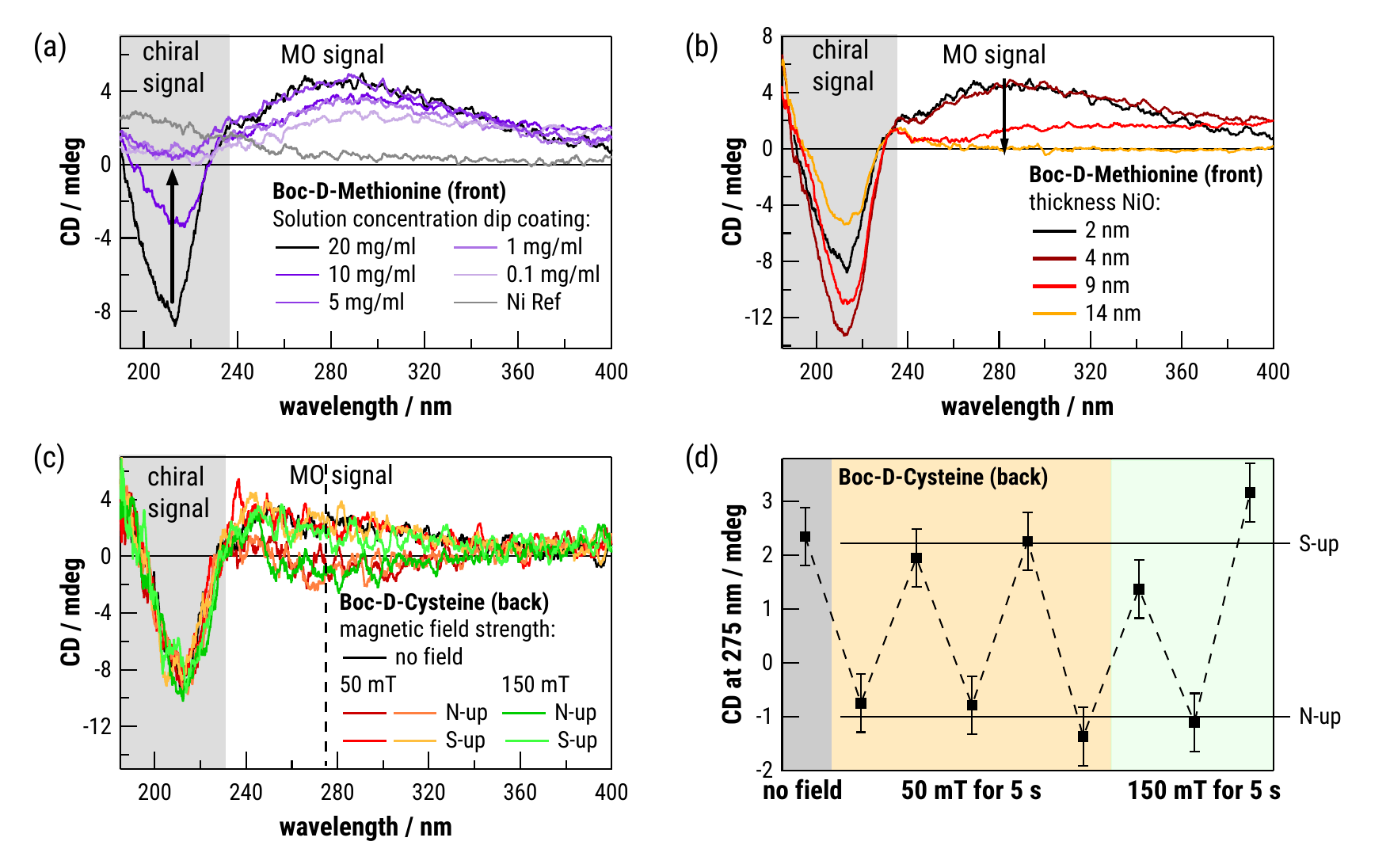}
    \caption{CD spectra in front configuration of Boc-D-methionine on Ni with (a) variation of the molecule concentration of the dipping solution and (b) variation of NiO layer thickness. (c) CD spectra in back configuration of Boc-D-cysteine on Ni after applying an opposing external magnetic field with a permanent magnet for 5\,s, line indicating wavelength where CD values in (d) are taken from; (d) CD values from (c) taken at 275\,nm with lines indicating average value signal strength while S-up and N-up denotes the orientation of the magnetic moments equivalent to MCD.}
    \label{Fig. Collection Magn Conc}
\end{figure}
Having confirmed the existence of an adsorption-induced chirality-dependent magneto-optical response, we next sought to determine its spatial origin and magnetic character. A first indication that the effect is confined to the interface is provided by its relatively small magnitude when compared to the MCD of an out-of-plane saturated pristine Ni film. Additionally, all measurements were performed in the absence of an external magnetic field, excluding field-induced magnetization as its origin. Instead, the observed signal reflects a remanent magnetic state generated upon molecular adsorption and localized within a thin interfacial region.

To verify the interface hypothesis the influence of molecule layer thickness on the adsorption-induced MO response was investigated. Therefore, we reduced the concentration of the dipping solution, thereby continuously lowering the amount of molecules being deposited. The reduction of the number of molecules absorbed on the surface of the NiO was tracked by the decrease in amplitude of the amino acid peak centered at 210\,nm (see arrow in Fig.~\ref{Fig. Collection Magn Conc}(a)). At the same time, the intensity of the adsorption-induced MO feature did not change thus proving independent of the molecular coverage. This indicates that only molecules directly in contact with the surface of the NiO layer contribute significantly to the generation of the observed MO signal centered around 275\,nm. This way the first molecular layer — and therefore the molecule–NiO/Ni interface — can be identified as the active region.

Further insight into the origin of the MO signal is obtained by comparing its spectral shape with the MCD spectrum of the pristine NiO/Ni film in the range between 190 and 260\,nm (Fig.~\ref{Fig. SPIPAC CD Methionine}(a)). The two spectra display remarkably similar line shapes, indicating that they arise from closely related electronic transitions. The adsorption-induced feature is, however, red-shifted by approximately 50\,nm closer resembling the bandgap of NiO with 3.6 to 4\,eV (350-310\,nm) \cite{shi2021temperature} (Fig.~\ref{Fig. SPIPAC CD Methionine}(c) and (e)). This demonstrates that molecular adsorption modifies the electronic structure associated with the magneto-optical transition rather than introducing an entirely new optical excitation. This behavior is consistent with electronic reconstruction at the molecule–NiO/Ni interface, where adsorption perturbs the local electronic environment while preserving the characteristic magneto-optical fingerprint of the substrate.

To further identify the active interfacial region, the thickness of the native oxide layer was systematically varied prior to molecular adsorption of Boc-D-Met (front configuration in Fig.~\ref{Fig. Collection Magn Conc}(b), see Fig.~S7(a) and (b) for both enantiomers and configurations). Increasing the oxide thickness progressively suppresses the adsorption-induced signal, which nearly disappears once the oxide exceeds approximately 4\,nm. This pronounced dependence demonstrates that the ultrathin oxide layer is the key player in the observed adsorption-induced signal rather than the underlying ferromagnetic Ni. Bulk NiO is an antiferromagnet with a N\'eel temperature well above room temperature, whereas ultrathin oxide layers may exhibit modified magnetic behavior arising from finite-size effects, reduced coordination, and uncompensated surface spins \cite{xu2019imaging,brand2025defect,proenca2011size}. Given the ultrathin NiO layer, the oxide likely resides in a paramagnetic state \cite{xu2019imaging}. The randomly oriented uncompensated spins at the surface of paramagnetic NiO would be flexible enough to reorient out-of-plane induced by the adsorption of the molecules (Fig.~\ref{Fig. SPIPAC CD Methionine}(g)). Although the present measurements do not directly resolve the magnetic structure of the oxide, the systematic suppression of the signal with increasing oxide thickness as well as its low thickness in the pristine NiO/Ni films strongly indicate that the active region is confined to the ultrathin molecule–NiO/Ni interface rather than the bulk metallic Ni or NiO.

A defining characteristic of an interfacial magnetic state is its response to an opposing external magnetic field. We therefore investigated the stability of the adsorption-induced signal by exposing the sample to an out-of-plane external magnetic field of 50\,mT for a duration of 5\,s between two CD measurements (Fig.~\ref{Fig. Collection Magn Conc}(c) and (d)). The orientation of the external field follows the notation of the one for MCD schematically shown in Fig.~\ref{Fig. SPIPAC CD Methionine}(b). The adsorption-induced MO signal follows the applied external field by inverting its sign but without any change in absolute signal strength. This reversal is stable upon several repetitions of the same and even a stronger external magnetic field of 150\,mT (Fig.~\ref{Fig. Collection Magn Conc}(d)). Here, after the magnetization, the induced response remains unchanged upon removal of the field, demonstrating that the interfacial state is remanent rather than field induced. Reversing the magnetic field reverses the sign of the induced signal, showing that the interfacial magnetic state can be reproducibly switched between two stable configurations. Moreover, the switching behavior closely follows the field dependence of the MCD response of the pristine NiO/Ni film in the low wavelength region, confirming that the adsorption-induced signal possesses a genuine magnetic character while remaining localized at the interface.

Taken together, these control experiments draw a coherent physical picture. The adsorption-induced response is absent on Au substrates, appears independently of molecular coverage thickness, retains the spectral characteristics of a magneto-optical transition, depends critically on the thickness of the native oxide layer, and remains reversibly switchable in the remanent state. Collectively, these observations demonstrate that adsorption of chiral molecules generates a chirality-dependent interface component with magnetic out-of-plane orientation localized at the molecule–NiO/Ni interface. Rather than interpreting this phenomenon solely within the framework of magnetization induced by proximity of adsorbed chiral molecules (MIPAC), we propose the broader concept of chiral spinterface formation, in which molecular adsorption reconstructs the electronic structure of the molecule–transitionmetaloxide interface, giving rise to a chirality-dependent interfacial magnetic response. Having established the existence, localization, and magnetic character of this interfacial state, we next turn to first-principles calculations to investigate its microscopic origin.

\subsection{Microscopic Origin of Chiral Spinterface Formation}
\begin{figure}[!ht]
    \centering
    \includegraphics[width=\linewidth]{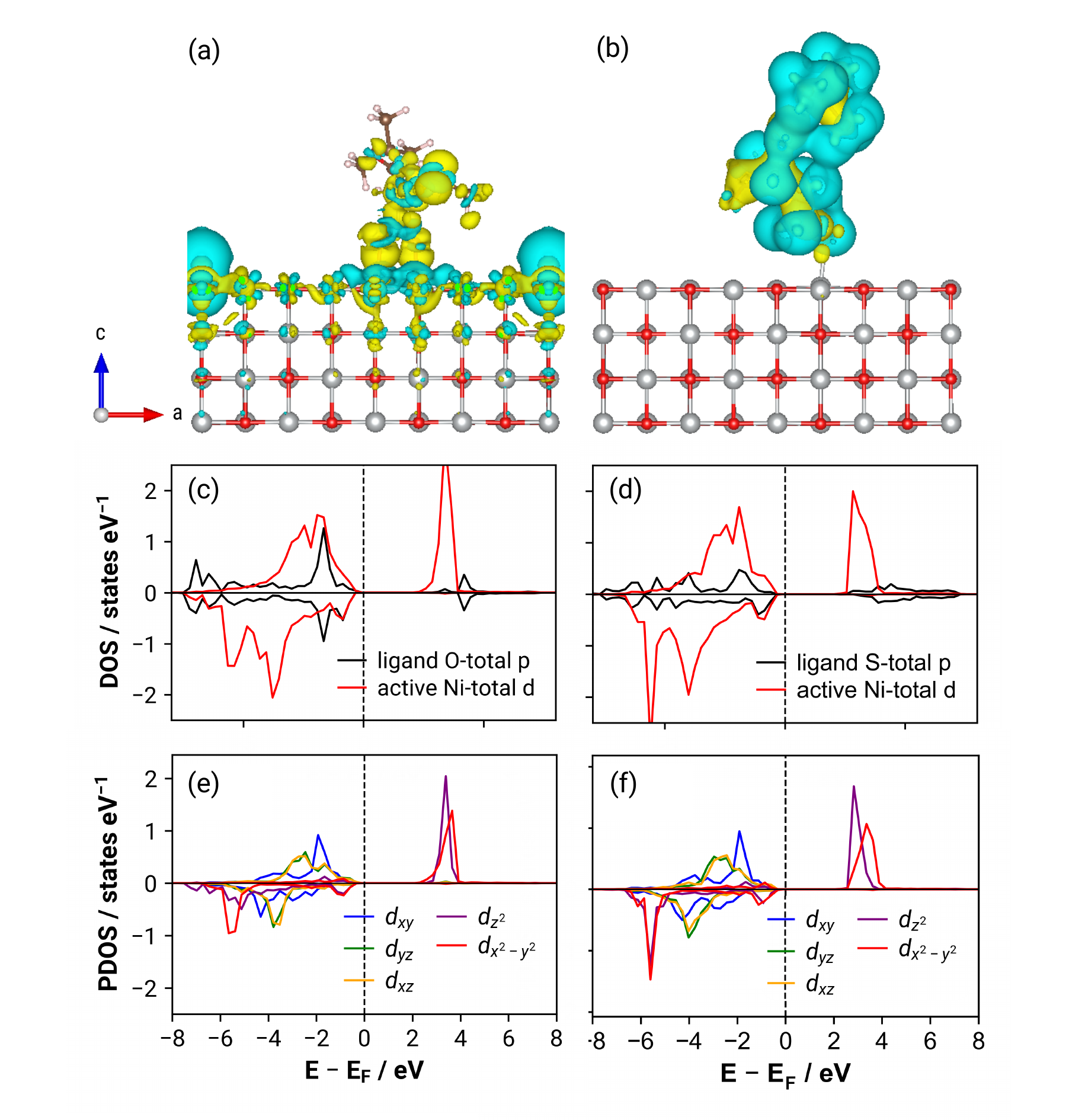}
    \caption{(a,b) Charge-density-difference maps for Boc-D-methionine adsorbed on NiO(100) through (a) the carboxylate group via two oxygen atoms and (b) the thioether sulfur atom. Yellow and blue regions indicate charge accumulation and charge depletion, respectively. (c,d) Projected density of states (PDOS) of the coordinating ligand and the surface Ni site for the O- and S-coordinated configurations, respectively. (e,f) Orbital-resolved Ni \textit{3d} PDOS for the corresponding O- and S-coordinated configurations. The dashed vertical line denotes the Fermi level.}
    \label{Fig. Adsorption geometries}
\end{figure}
To understand how chiral molecular adsorption modifies the NiO interface, we performed density functional theory (DFT) calculations of Boc-L- and Boc-D-methionine on NiO(100). This surface was selected as a representative model because it is the most stable low-index facet of stoichiometric rocksalt NiO \cite{xiang2018nio,hermawan2020octahedral,oliver1993computer}, providing a well-defined reference for investigating molecule–oxide interactions.

Two adsorption configurations were considered (Fig.~\ref{Fig. Adsorption geometries}): bidentate coordination through the deprotonated carboxylate group (Fig.~\ref{Fig. Adsorption geometries}(a)) and binding through the thioether sulfur atom (Fig.~\ref{Fig. Adsorption geometries}(b)). Carboxylate binding is energetically favored for both enantiomers, with adsorption energies of -0.91\,eV, compared with -0.36\,eV (Boc-D-Met) and -0.34\,eV (Boc-L-Met) for sulfur coordination. The nearly identical adsorption energies of the two enantiomers indicate no significant energetic preference for either handedness on the ideal NiO(100) surface.

To examine the electronic interactions underlying these adsorption configurations, we analyzed the charge-density difference (CDD), defined as
\begin{equation}
    \Delta \rho = \rho (\text{Boc-Met/NiO})-\rho (\text{NiO})-\rho (\text{Boc-Met}). 
\end{equation}
For carboxylate coordination, the molecule binds through two oxygen atoms to surface Ni sites, while the dissociated proton remains on the surface to maintain charge neutrality. The CDD maps reveal pronounced charge redistribution around the interfacial Ni-O bonds, extending into the subsurface NiO layers (Fig.~\ref{Fig. Adsorption geometries}(a)). In contrast, sulfur coordination produces predominantly intramolecular charge redistribution, with a much weaker perturbation of the oxide surface (Fig.~\ref{Fig. Adsorption geometries}(b)). These results indicate substantially stronger electronic coupling at the carboxylate-bound interface.

The projected density of states (PDOS) further supports this picture. For carboxylate coordination, the O \textit{2p} and surface Ni \textit{3d} states exhibit substantial spectral overlap between -2 and -1\,eV (Fig.~\ref{Fig. Adsorption geometries}(c)), consistent with interfacial orbital hybridization. The sulfur-bound configuration shows weaker S \textit{3p}-Ni \textit{3d} overlap (Fig.~\ref{Fig. Adsorption geometries}(d)), reflecting the reduced molecule-surface interaction. Orbital-resolved Ni \textit{3d} PDOS reveals pronounced changes in the $d_{z^2}$ and $d_{x^2-y^2}$ states upon carboxylate adsorption, with smaller modifications in the remaining \textit{3d} orbitals (Fig.~\ref{Fig. Adsorption geometries}(e) and (f)). These changes reflect the altered local electronic environment of surface Ni atoms, whereas sulfur coordination produces only minor changes in the Ni \textit{3d} electronic structure.

Together, the adsorption energetics, charge-density redistribution, and PDOS analyses establish carboxylate coordination as the preferred binding motif, accompanied by substantial modification of the NiO surface electronic structure. Although both enantiomers exhibit nearly identical adsorption energetics, their strong electronic coupling to the magnetic oxide provides a possible pathway for chirality-dependent interfacial interactions. These findings suggest that the experimentally observed magneto-optical response may involve the interplay between molecular chirality and the electronic and magnetic properties of the interface, rather than differences in adsorption stability.

\section{Conclusion}\label{sec13}
We demonstrate that adsorption of chiral molecules on NiO/Ni thin films induces a remanent, chirality-dependent magneto-optical response localized at the molecule-oxide interface. Systematic experiments establish its interfacial origin and magnetic character, while first-principles calculations reveal how molecular adsorption modifies the electronic structure of the NiO surface through interfacial bonding, charge redistribution, and orbital hybridization.

These findings identify \textbf{chiral spinterfaces as an important, previously overlooked component of CISS-related systems}, challenging the conventional view of molecule-electrode interfaces as passive contacts. Our results highlight the active role of molecular interfaces in chirality-dependent magnetic phenomena and open new opportunities for understanding and engineering spin functionality through molecular adsorption and interface design.

\section{Methods}\label{sec11}
\subsection{Sample Preparation}
The 30\,nm Nickel thin films were deposited via electron beam evaporation with a base pressure of \SI{4e-7}{\milli\bar} on a 650\,\textmu m thick c-plane sapphire substrate from \textit{Siegert Wafer}. The wafer was separated into $(8\times 5)$\,\si{\milli\meter\squared} sized sample pieces. Before the molecule deposition, the samples were ultrasonicated in isopropanol for 10\,min to reduce harsh contamination. Residual organic contamination was removed by plasma etching. The etching was performed in a Zepto low-pressure plasma etching chamber by \textit{diener electronics} with 100\,\% power and 5\,mbar Argon gas pressure for 1 minute. Following this was the direct transfer of the clean substrates into the molecule solution. The molecules were dissolved in spectroscopic-grade ethanol with a concentration of \SI{20}{\milli\gram\per\milli\litre} if not stated otherwise.\\
N-(tert-butoxyxarbonyl)-L-methionine (99\,\%), N-(tert-butoxyxarbonyl)-D-methionine (98\,\%), N-(tert-butoxyxarbonyl)-L-cysteine (99\,\%), and $\alpha$-L-polypeptide C[AAAAK]$_6$ were ordered from \textit{Sigma-Aldrich}, while N-(tert-butoxyxarbonyl)-D-cysteine (98\,\%) from \textit{Iris Biotech}. The synthesis of Bi$_{38}$O$_{45}$(Boc-L-Met)$_{24}$ was carried out according to the procedure in the literature \cite{morgenstern2024anchoring} and Bi$_{38}$O$_{45}$(Boc-D--Met)$_{24}$ according to \cite{hornig2026ligand}.
\subsection{Circular Dichroism Spectroscopy}
The CD measurements were performed with a J-1500 CD spectrophotometer from \textit{JASCO}
equipped with a self designed rotation sample stage described in Ref. \cite{schoelzel2026stepbystepworkflowextractgenuine}. For all thin film measurements the azimuthal angle was varied between 0\,$^\circ$ and 180\,$^\circ$ in 15\,$^\circ$ steps with a spectrum taken at each angle. All shown spectra represent the average over all 13 respective measurements. The CD spectra were collected at a scanning speed of 100 nm/min with a step size of 0.1 nm, a digital integration time (D.I.T.) of 2 s and a bandwidth of 1 nm.\\
For the MCD measurements, the J-1500 was either equipped with the PMCD-586 permanent magnet with a field strength of 1.6 T or a self-designed permanent magnet holder, which allows for a field variation by changing the distance between the magnets. The spectra were collected with the same parameters as the thin films.
\subsection{X-Ray Reflectivity/diffraction}
X-Ray reflectivity (XRR) and X-Ray Diffraction (XRD) were performed using a rotating-anode \textit{Rigaku} SmartLab diffractometer with a 5-axis goniometer. XRR was measured using a K\textbeta~filter to balance beam intensity and spectral resolution. The resulting XRR curves were fitted using GenX3, version 3.6.27 to extract layer thicknesses, interface roughnesses, and the scattering length density (SLD) depth profile. The density profile was modeled using the following stack: Al$_2$O$_3$(substrate) / Ni / NiOx.\\
Symmetric out-of-plane XRD (OOP-XRD) was measured using a two-bounce Ge(220) monochromator, selecting the Cu K$_{\alpha,1}$ line, for the annealed samples and with Bragg-Bretano geometry and K\textbeta~filter for the reference sample.\\
Pole figures were performed using the IP arm option of the Rigaku Smartlab.
\subsection{X-Ray photoelectron spectroscopy}
X-ray photoelectron spectroscopy (XPS) was performed with an ESCALAB 250Xi photoelectron spectrometer from \textit{ThermoFisher Scientific} in an ultra-high vacuum (UHV) chamber utilizing a monochromatic Al-K$_\alpha$ (1486.68\,eV) X-ray source with a beam diameter of 650\,\textmu m. During measurements the base pressure was better than $2\times10^{-9}$\,mbar. The binding energies of all spectra were referenced to the binding energy of the C 1\textit{s} $(285\pm0.1)$\,eV. All core level spectra were collected with a pass energy of 20\,eV.
\subsection{Superconducting Quantum Interference Device Vibrating-Sample Magnetometry}
The magnetic hysteresis loops, $M(H)$, were measured using an MPMS3 SQUID-VSM device (\textit{Quantum Design}). The measurements were performed at room temperature in the out-of-plane (oop) and in-plane (ip) geometries. For the oop geometry the sample was placed in a lab-fabricated plastic straw and the magnetic field was applied perpendicular to the film plane. For the ip geometry the sample was placed on the standard quartz holder of the device and the magnetic field was applied parallel to the film plane.
\subsection{Computational Methods}
Spin-polarized density functional theory (DFT) calculations were performed using the Vienna Ab initio Simulation Package (VASP) \cite{kresse1996efficient}. The exchange-correlation energy was described within the generalized gradient approximation using the Perdew-Burke-Ernzerhof (PBE) \cite{perdew1996generalized} functional. To account for the localized nature of the Ni 3\textit{d} electrons, the DFT+U method was employed in the Dudarev formalism \cite{dudarev1998electron} using an effective Hubbard parameter of $U_\text{eff}$ = 5.3\,eV for Ni (corresponding to $U$ = 6.3\,eV and $J$ = 1.0\,eV). The interaction between the ionic cores and valence electrons was described using the projector augmented-wave (PAW) \cite{kresse1999ultrasoft} method, and the Kohn–Sham wavefunctions were expanded in a plane-wave basis with a kinetic energy cutoff of 500\,eV.\\
The NiO(100) surface was modeled using an antiferromagnetic 4$\times$4 slab, preserving the magnetic ground-state ordering of bulk NiO, with lattice parameters of a = b = \SI{16.59}{\angstrom} and c = \SI{34.28}{\angstrom}. A vacuum layer of \SI{15}{\angstrom} was introduced along the surface normal to eliminate interactions between periodic images. The slab model and computational settings for the clean NiO(100) surface were adopted from the work of Apergi~\etal \cite{apergi2020tuning}. Brillouin-zone integrations were performed using a \textGamma-centered $2\times 2 \times 1$ Monkhorst-Pack k-point mesh.\\
All atomic structures were fully relaxed using the conjugate-gradient algorithm until the total energy converged to $10^{-5}$\,eV and the residual forces on every atom were below \SI{0.01}{\eV\per\angstrom}. Symmetry constraints were removed during structural optimization to allow unrestricted relaxation of the adsorbate–surface system. Boc-Methionine adsorption was investigated for both L- and D-enantiomers in sulfur-coordinated and carboxyl-coordinated adsorption configurations. 

\bibliography{library}
\backmatter

\bmhead{Supplementary information}

See supplementary material for additional information on the initial characterization and the control experiments.

\bmhead{Acknowledgements}

F. S., D. H., L. R., A. K., C. L., M. M., O. H. and G. S. gratefully acknowledge funding through German Research Foundation (DFG), TRR 386, TP (B3, A4, Z2), project number 514664767. The authors thank the Center for Micro and Nanotechnologies (TU Chemnitz) for the deposition of the Ni films. A.G. and S. T. would like to thank Geert Brocks for discussions of the DFT calculations.

\bmhead{Contributions}
F.S. designed and carried out the sample preparation, CD and XPS measurements under the supervision of G.S. R.E. performed and analyzed all XRD and XRR measurements, while L.R. conducted the SQUID-VSM measurements, both under the supervision of O.H. D.H. synthesized the BiO-NCs under the supervision of M.M. A.G. performed the first-principles calculations under the supervision of S.T. A.L., J.L., A.K. and C.L. contributed to the interpretation of the experimental results through scientific discussions. F.S., A.G., G.S. and S.T. wrote the manuscript. All authors discussed the results and contributed to the manuscript and Supplementary Information.

\end{document}


\setlength{\tabcolsep}{12pt}
\supplementarysection

\section{Initial Characterization of the Ni thin films}\label{secA1}
\begin{figure}[!ht]
    \centering
    \includegraphics[width=0.9\linewidth]{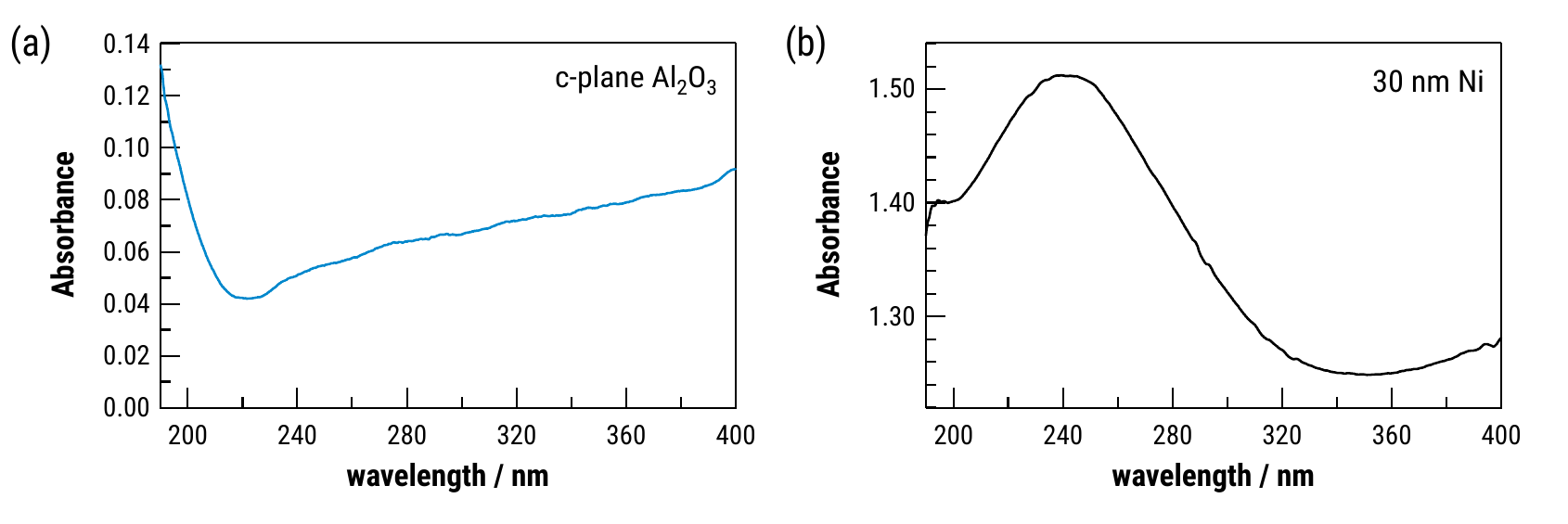}
    \caption{Absorbance spectra of (a) c-plane sapphire and (b) 30\,nm Ni with sapphire substrate subtracted. Both show a low transmission making the substrate suitable for CD measurements in transmission geometry.}
    \label{SI. Absorbance}
\end{figure}
\noindent
The following section consists of additional experimental data on the initial characterization of the Ni thin films on sapphire used as substrates in all experiments. These experiments aim to establish a well-defined platform for the structural, optical and magnetic properties to reliably determine changes induced by molecule adsorption.

In Fig.~\ref{SI. Absorbance} the absorbance spectra of the pure sapphire substrate (Fig.~\ref{SI. Absorbance}(a)) and the 30\,nm Ni film are displayed (Fig.~\ref{SI. Absorbance}(b)). The absorbance of the c-plane sapphire substrate (Fig.~\ref{SI. Absorbance}(a)) is close to zero down to the deep UV-range (185\,nm). Due to this low absorbance sapphire is the ideal substrate for the transmission experiments. Although the Ni film absorbs two orders of magnitude more light (Fig.~\ref{SI. Absorbance}(b)), its transmission is still sufficient to enable the CD measurements. All spectra were collected with the J-1500 spectrophotometer simultaneously to the CD measurements with the parameters described in the methods section. For the c-plane sapphire a measurement of the empty sample holder was subtracted while the sapphire measurement acted as baseline for the Ni films thus ensuring that any absorption originated purely from the Ni.

\begin{figure}[h]
    \centering
    \includegraphics[width=\linewidth]{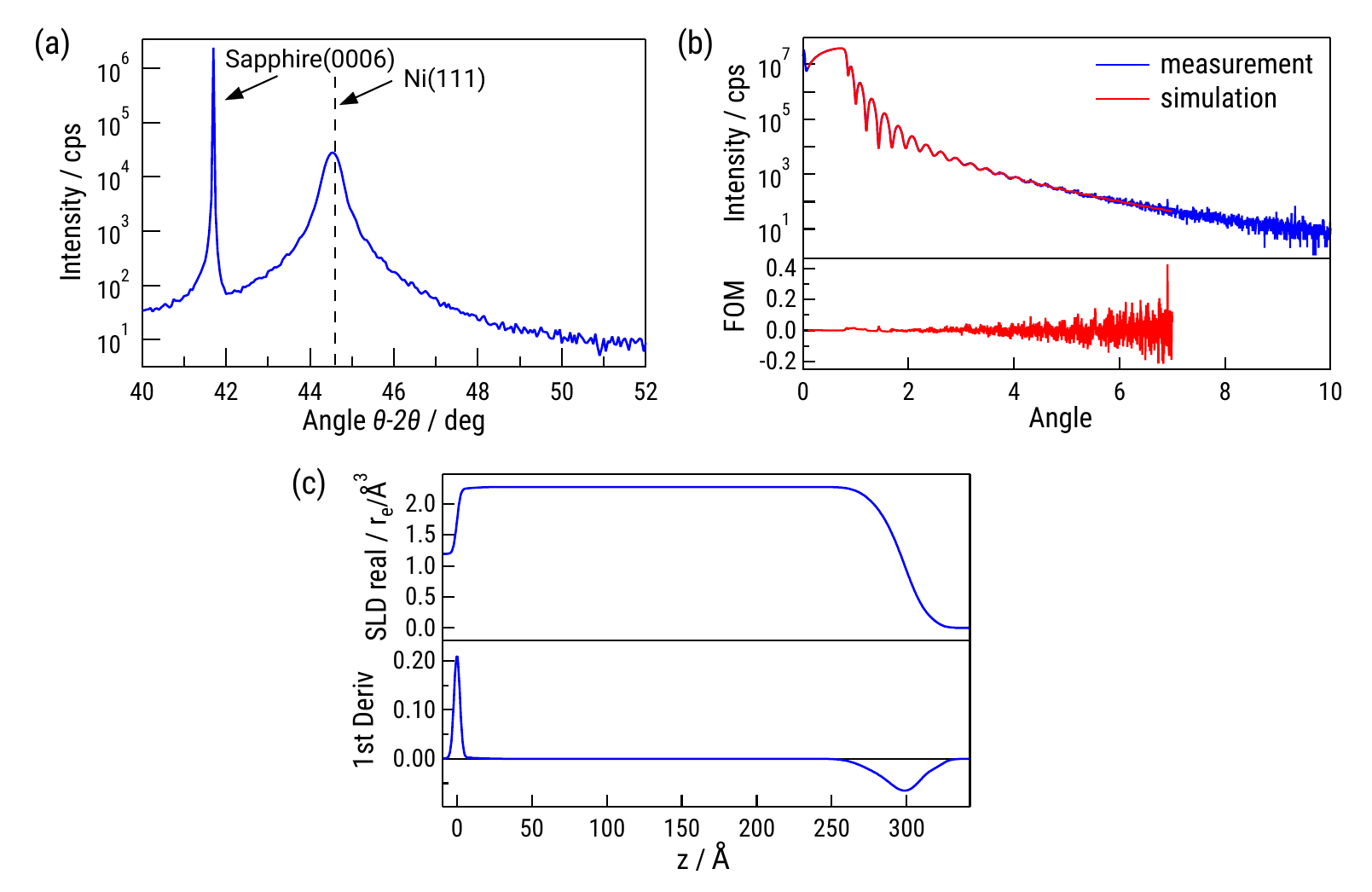}
    \caption{(a) XRD $\theta-2\theta$ scans of the as deposited reference Ni sample in Bragg Brentano geometry with a K\textbeta~filter. (b) XRR measurement of the same sample with the best fit utilizing a simple Al$_2$O$_3$/Ni/NiO$_\text{X}$ model. (c) Real part of the SLD and its first derivative as extracted from the best fit.}
    \label{SI. XRD XRR}
\end{figure}

\begin{figure}[H]
    \centering
    \includegraphics[width=0.5\linewidth]{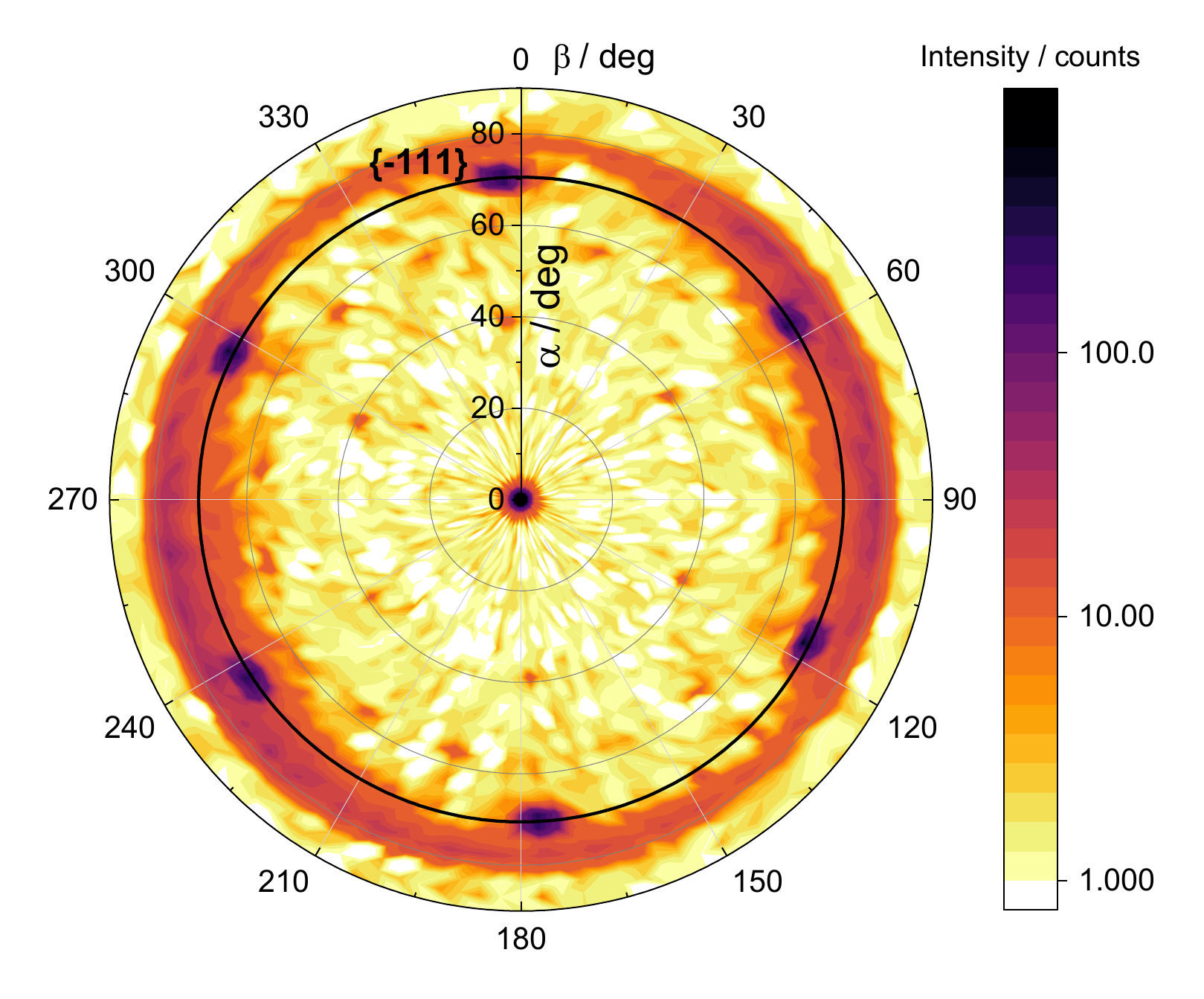}
    \caption{Pole figure recorded for the Ni(111) reflex at 2$\theta=41.56\,^\circ$ for the 30\,nm Ni substrate. $\alpha$ denotes the polar angle relative to the film normal and $\beta$ is the azimuth angle of the ip rotation. The black reference line corresponds to the polar angle of the $\{-111\}$ lattice plane family.}
    \label{SI. Pole figure}
\end{figure}

\begin{figure}[!h]
    \centering
    \includegraphics[width=0.9\linewidth]{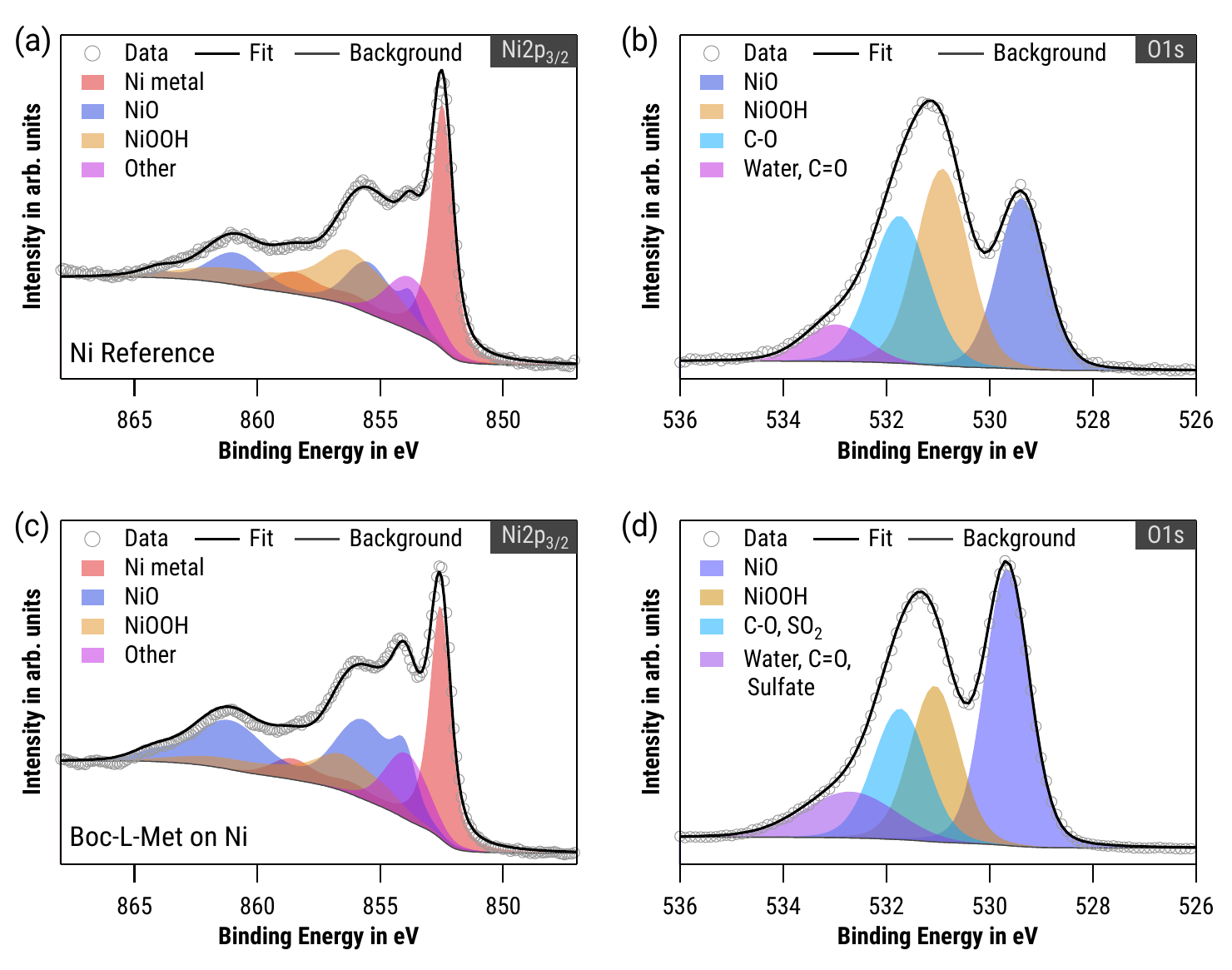}
    \caption{Deconvoluted (a)/(c) Ni2$p_{1/2}$ and (b)/(d) O1$s$ XPS core level spectra of the pure Ni substrate (a-b) and Boc-L-methionine on Ni (c-d). Both core levels share signals from metallic Ni (Ni metal) }
    \label{SI. XPS pure Ni}
\end{figure}

For the structural characterization of the Ni thin films we conducted both XRD as well as XRR measurements with a Rigaku SmartLab diffractometer. In the XRD $\theta-2\theta$ scans (Fig.~\ref{SI. XRD XRR}(a)) of the as deposited reference Ni sample a well defined Ni(111) reflex is visible, pointing towards epitaxial growth. Additionally, the double peak of Sapphire(0006) caused by the K$_{\alpha,1}$ and K$_{\alpha,2}$ lines is observed. The XRR measurement was used to determine the thickness of both the Ni film and the native oxide layer. We fitted the data with a simple Al$_2$O$_3$/Ni/NiO$_\text{X}$ model, where the substrate and Ni density was fixed to its bulk value (Fig.~\ref{SI. XRD XRR}(b)). Meanwhile the NiO$_\text{X}$ density was being fitted, as well as the thicknesses, roughnesses and instrument parameters. From the  real part of the resulting scattering length density (SLD) (Fig.~\ref{SI. XRD XRR}(c)) the Ni thickness was determined to 27.3\,nm and the one of NiO to around 2\,nm. The combination of Ni roughness and NiO$_\text{X}$ layer caused the broad asymmetric peak in the first derivative at the surface.
To access the in-plane (ip) structure of the Ni films a pole figure was recorded for the Ni(111) reflex at 2$\theta=41.56\,^\circ$ (Fig.~\ref{SI. Pole figure}). For a single crystal, 3 spots are expected for the (-111), (1-11) and (11-1) lattice plane, respectively. The coexistence of grains with ABC and CBA stacking directions creates another set of 3 reflexes, which creates a 6-fold symmetry, which is consistent with the dark spots in the figure.
XPS was utilized to determine the chemical composition of the native oxide layer on top of the bare Ni substrate. Therefore both the Ni 2\textit{p} as well as the O 1\textit{s} core levels were measured and deconvoluted (Fig.~\ref{SI. XPS pure Ni}(a-b)). A Shirley background was applied before fitting for both core levels \cite{shirley1972high}. The Ni2$p_{3/2}$ core level (Fig.~\ref{SI. XPS pure Ni}(a)) was fitted with a Mahan function for the metallic component (Ni metal) and Voigt functions for the oxides. Multiplett splitting was applied after Biesinger~\etal\cite{biesinger2011resolving}. Besides the metallic component at 852.4\,eV signals from both NiO (854.1\,eV) as well as NiOOH (854.6\,eV) can be observed. An additional other component is needed to achieve a reasonable fit. We suspect that the software underestimated the background at this position as no chemical state of the Ni could be identified from its binding energy position. Analogous components for the native oxide of Ni can also be found in the O 1\textit{s} core level spectrum (Fig.~\ref{SI. XPS pure Ni}(b)), which was fitted with Voigt functions. Both signals from NiO as well as NiOOH could be identified. Additionally residual water and organic contamination from single (C-O) and double bonds (C=O) to Carbon were observed.

To compare how the molecule deposition process affects the chemical composition analogous measurements of the Ni 2\textit{p} and O 1\textit{s} core levels were performed for Boc-L-methionine (Fig.~\ref{SI. XPS pure Ni}(c-d)). In the Ni 2\textit{p} detail spectrum an increased intensity ratio of NiO/NiOOH can be observed. This can be explained by a reduced amount of water on the substrate surface resulted from the plasma etching treatment, followed by the direct transition to the molecule solution. Here, the molecules cover the surface shielding the NiO surface from water and therefore reducing the signal from this component. This is mirrored in the O 1\textit{s} where the NiOOH component has less intensity compared to the NiO. As water and C=O cannot be distinguished in the O 1\textit{s} detail spectrum and Boc-L-Met exhibits two carbon oxide double bonds a reduction of water on the surface cannot be determined. Due to a similar reason we cannot determine if the carboxylate motif is dominant against the sulfur absorption. As Ni atoms bound to Oxygen in the NiO layer display a chemical bond similar to the one expected for the polar Ni-O bonds formed during absorption of Boc-Met they cannot be distinguished in their binding energy.

\begin{figure}[!h]
    \centering
    \includegraphics[width=\linewidth]{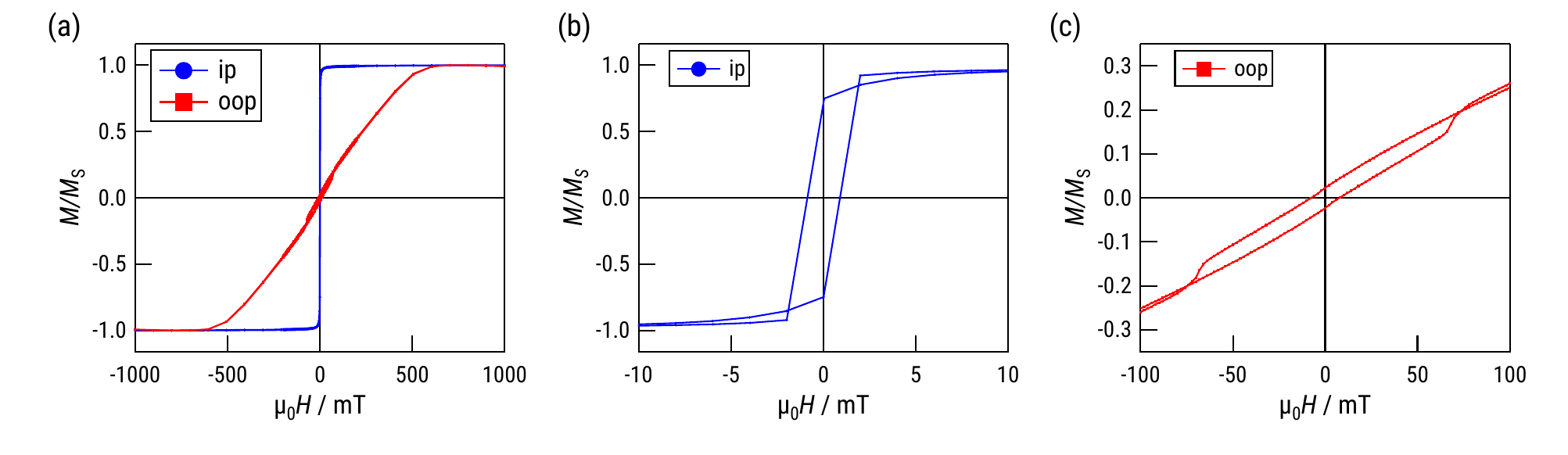}
    \caption{(a) SQUID-VSM measurements of the pure 30\,nm Ni film on c-plane sapphire in in-plane and out-of-plane geometry with zoom in around zero for (b) ip and (c) oop. A very pronounced ip easy axis with low coercive field is visible, with the oop direction being the hard axis.}
    \label{SI. SQUID Ni}
\end{figure}
SQUID-VSM measurements were conducted (Fig.~\ref{SI. SQUID Ni}) to determine the magnetic properties of the bare Ni films. Here, a very pronounced ip easy axis with low coercive field (3\,mT) is observed (Fig.~\ref{SI. SQUID Ni}(b)) with the oop direction being the hard axis (Fig.~\ref{SI. SQUID Ni}(c)). For a saturation in oop direction an external magnetic field of 600\,mT is needed (red curve in Fig.~\ref{SI. SQUID Ni}(a)).

\section{Additional Data for the adsorption-induced spinterface}
\begin{figure}[!h]
    \centering
    \includegraphics[width=0.95\linewidth]{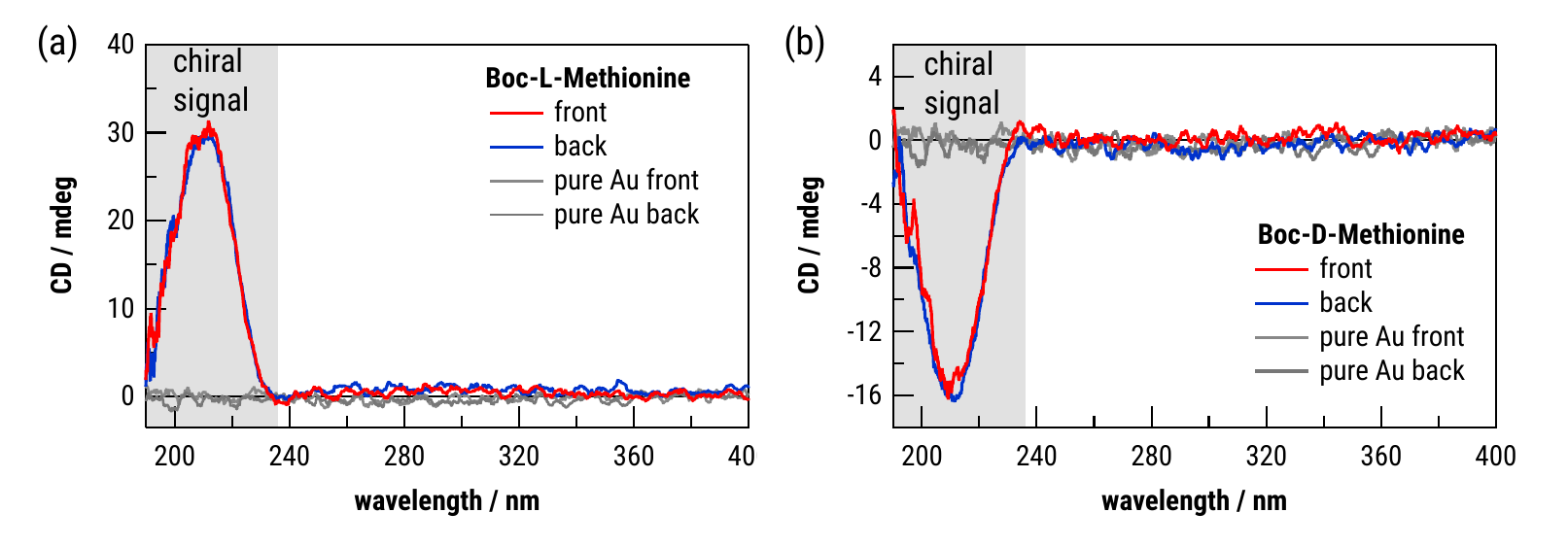}
    \caption{CD spectra in front and back configuration of (a) Boc-L-methionine and (b) Boc-D-methionine on 30\,nm Gold thin film on c-plane sapphire.}
    \label{SI. Met on Au}
\end{figure}

\begin{figure}[!h]
    \centering
    \includegraphics[width=0.95\linewidth]{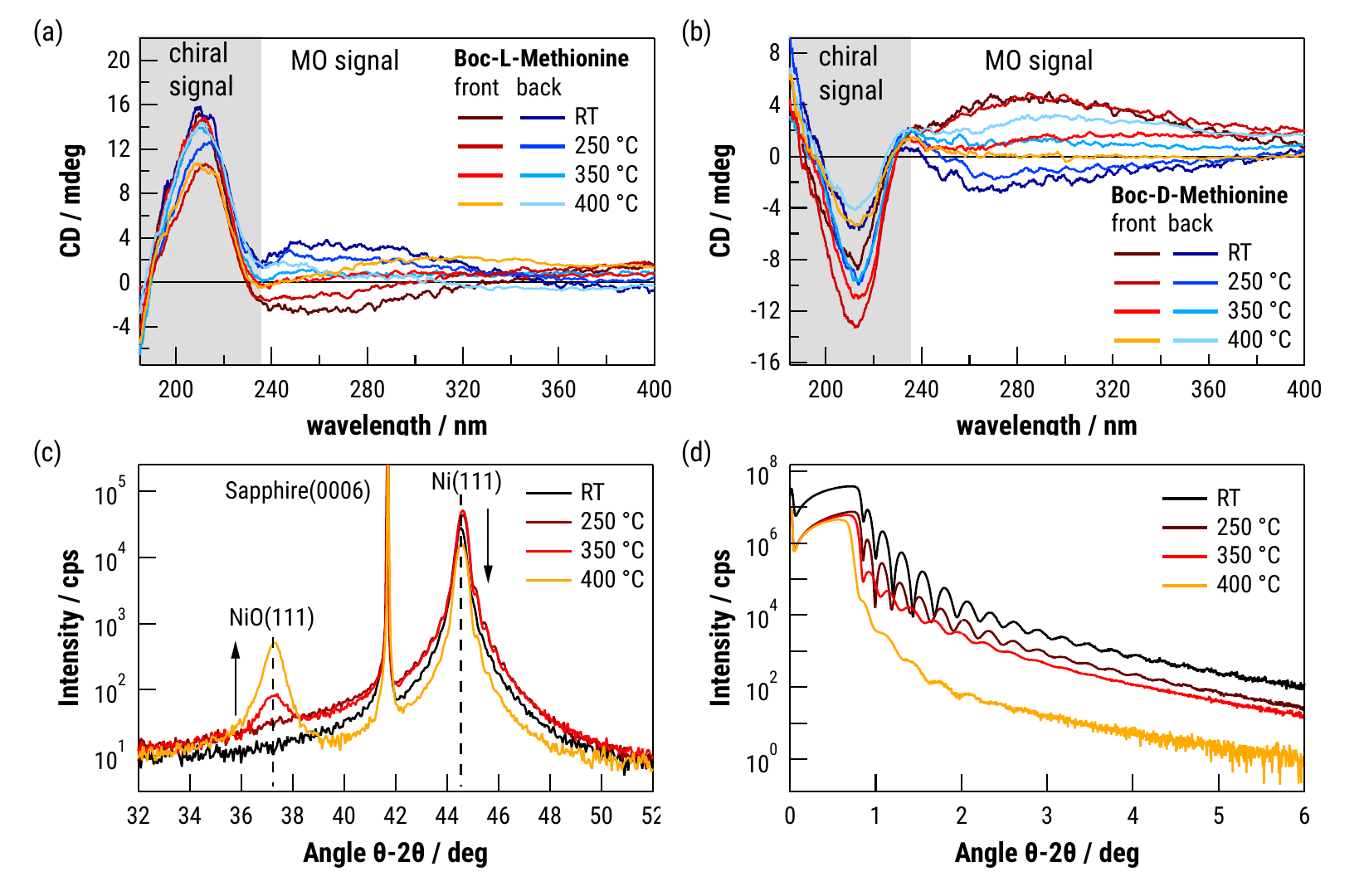}
    \caption{CD spectra in front and back configuration of (a) Boc-L- and (b) Boc-D-methionine on Ni films annealed at different temperatures resulting in an increase of NiO layer thickness with increasing temperature. (c) XRD $\theta-2\theta$ scans for the different samples and (d) XRR $\theta-2\theta$ scans showing in increase in roughness of the films. }
    \label{SI. NiO XRD XRR}
\end{figure}

In this section additional experimental data for the control experiments is supplied. For the first one both enantiomers of Boc-methionine were deposited on 30\,nm Au films following the exact same sample preparation protocol as the Ni films. The resulting CD spectra (Fig.~\ref{SI. Met on Au}) show the characteristic amino acid peak centered at 210\,nm while lacking the MO response observed for the magnetic films. Therefore, the origin of the adsorption-induced signal can be traced to the Ni film rather than the molecules.
The second control experiment investigates the influence of the NiO thickness on the signal strength of the MO response. A series of annealing processes were conducted resulting in a continuous increase in layer thickness. The bare Ni films were annealed in a LHM01 oven from \textit{SNOL} at ambient conditions. For parameters please refer to Tab.~\ref{Tab. NiO}. For temperatures higher than 250$\,^\circ$C a NiO(111) reflex can be observed in the XRD scans showing the epitaxial growth of Ni getting transferred to the NiO (Fig~\ref{SI. NiO XRD XRR}(c)).

The layer thickness of the NiO and Ni films were determined from XRR fits and from the NiO(111) reflex via the Scherrer equation. The thicknesses can be found in Tab.~\ref{Tab. NiO}. For the RT and 250$\,^\circ$C annealed samples no Scherrer equation could be applied due to the absence of the NiO(111) reflex (see Fig~\ref{SI. NiO XRD XRR}(c)). For the two higher temperatures the XRR fit was impossible to realize due to high surface roughness (Fig.~\ref{SI. NiO XRD XRR}(d)).

These samples were used for deposition of Boc-L- and Boc-D-methionine following our deposition protocol. The resulting CD spectra can be found in Fig.~\ref{SI. NiO XRD XRR}(a-b), where a MO signal can only be observed up to the 250$\,^\circ$C / 4\,nm NiO film. Above this annealing temperature/NiO thickness no adsorption-induced MO response is visible.

\begin{table}[!h]
    \caption{Estimated thicknesses and measured CD intensity of the adsorption-induced MO signal for Ni films after annealing at different temperatures. The thickness was both determined via XRR-fitting and calculated from XRD using the Scherrer equation.}
    \setlength{\tabcolsep}{3pt}
    \centering
    \begin{tabular}{@{}ccccccc@{}}
    \hline
    $T$ / $^\circ$C & $t$ / min & \multicolumn{2}{c}{$d_\text{Ni}$ / nm}& \multicolumn{2}{c}{$d_\text{NiO}$ / nm} & $CD$ at 275\,nm / mdeg\\\cmidrule{3-6}
     \multicolumn{2}{c}{ }& XRR & XRD & XRR & XRD & \\
     \hline
    RT     & -- & 27.3 $\pm$ 1.0 & 27.1 $\pm$ 1.0 & 2.0 $\pm$ 1.0 & -- & 6.02 $\pm$ 1.66 \\
    250     & 120 & 26.4 $\pm$ 1.0 & 28.6 $\pm$ 1.0 &3.8 $\pm$ 1.5 & -- & 2.12 $\pm$ 1.83 \\
    350     & 60 & 25.5 $\pm$ 1.0 & 25.0 $\pm$ 1.0 & 8 $\pm$ 1.5 & 9.2 $\pm$ 1.5 & -- \\
    400     & 90 & 23.1 $\pm$ 1.0 & 22.3 $\pm$ 1.0 & (20 $\pm$ 1.5) & 13.5 $\pm$ 1.5 & -- \\
    \hline
    \end{tabular}
    \label{Tab. NiO}
\end{table}

\bibliography{library}